\pdfoutput=1

\documentclass[galaxies,article,accept,moreauthors,pdftex]{Definitions/mdpi}

\usepackage{xcolor}
\usepackage{graphicx} 
\usepackage{tikz}
\usepackage{mathtools}
\DeclarePairedDelimiter\abs{\lvert}{\rvert}%

\firstpage{1} 
\pubvolume{1}
\issuenum{1}
\articlenumber{0}
\pubyear{2026}
\copyrightyear{2026}
\externaleditor{Firstname Lastname} 
\datereceived{30 April 2026 } 
\daterevised{19 August 2026 } 
\dateaccepted{9 September 2026 } 
\datepublished{ } 
\pdfoutput=1 

\Title{Neutrino 
 Emission from Proton--Photon Jet Interactions in~Microquasars}

\Author{Theodoros Smponias 
}
\AuthorNames{Theodoros Smponias}

\address[1]{Directorate of Primary Education of the Ionian Islands, 
 49100 Corfu, Greece; t.smponias@hushmail.com}

\abstract{
Microquasars are candidate sites for high-energy particle production within our galaxy. Because their distances and compact emission regions limit direct observational constraints, numerical simulations are useful for connecting jet dynamics with possible multi-messenger signatures. This work models neutrino production from relativistic magneto-hydrodynamic microquasar jets, focusing on the proton--photon ($p\gamma$) channel associated with interactions between accelerated protons and ambient photon fields. Proton--proton ($pp$) interactions are acknowledged as a possible process in dense environments, but they are not modeled in the present calculation. A ray-tracing procedure is applied to the hydrodynamic simulation output to produce synthetic neutrino images from a stationary observer's perspective. Synthetic spectra and intensity maps are presented and compared with representative sensitivities of current and future neutrino detectors.
}


\begin{document}

\section{Introduction}
\label{intro}

Microquasars (MQs) are binary stellar systems comprising a main sequence star orbiting a collapsed stellar remnant \cite{Mirabel99}. Mass transfer onto the compact object powers relativistic jets launched largely perpendicular to the orbital plane. These jets emit radiation across the electromagnetic spectrum, from radio to very-high-energy (VHE) gamma rays, and may also produce neutrinos under hadronic interaction scenarios \cite{Romero2003,Reynoso2008,Reynoso2009,Reynoso2019}. Observations such as apparent superluminal motion support bulk hadronic flows in these jets \cite{Romero2003}. Microquasar neutrino and hadronic emission scenarios have also been discussed in earlier work on X-ray binaries and specific MQ systems \cite{Christiansen2006,Romero2005LSI,Christiansen2013Jets}.

Recent ultra-high-energy (UHE) gamma-ray observations strengthen the case for galactic compact-object jet systems as potential PeVatron accelerators. In particular, results from LHAASO provide a new UHE view of galactic gamma-ray sources and motivate renewed interest in jet-powered systems as extreme particle accelerators \cite{Cao2024LHAASOcat}. Furthermore, UHE gamma-ray emission associated with black-hole jet systems has been discussed in the context of LHAASO detections, providing additional motivation for hadronic acceleration scenarios and associated neutrino production \cite{LHAASO2024BHjets}. In parallel, microquasar jet and jet--environment interaction models have recently been explored as PeVatron candidates, including jet--cocoon systems and microquasar remnants as potentially ``hidden'' PeVatrons~\cite{Wang2025,Zhang2025,Abaroa2025}.


This work does not aim to reintroduce the general framework of neutrino emission from microquasar jets; related scenarios have been discussed in earlier work, including the author's previous Galaxies papers. The present manuscript focuses on (i) the proton--photon ($p\gamma$) component in a RMHD-based jet pipeline, (ii) the numerical setup and assumptions used for acceleration, diffusion, magnetic fields, and target photon fields, and (iii) a simple distance-dependent detectability estimate derived from published instrument sensitivities. Proton--proton ($pp$) interactions are relevant in dense target environments and are mentioned for context, but they are not included in the emission calculation presented~here.

These developments motivate time-dependent, simulation-based modeling that links RMHD jet dynamics and interaction sites (shocks, blob fronts, terminal regions) to neutrino observables. Since MQ distances and complex geometries limit direct inference, detailed numerical modeling is essential to connect theory with multi-messenger observables.

Strong magnetic fields near jet bases and their tangled structures justify a fluid description via special relativistic magneto-hydrodynamics (RMHD) \cite{Koessl1990,Rieger2006,Rieger2019,Singh2019MHD}. Toroidal magnetic components collimate the jet \cite{Koessl1990,Singh2019MHD}, while interactions with stellar and disk winds shape jet morphology and confinement \cite{Hughes1991,Reynoso2009}.

In this work, we extend previous modeling efforts by coupling relativistic magnetohydrodynamic (RMHD) simulations of microquasar jets with a neutrino-emission pipeline for the proton--photon ($p\gamma$) channel. Special emphasis is placed on transient structures within the jet, such as overdense blobs and shock regions, where accelerated protons interact with ambient photon fields.

Our study therefore combines RMHD jet evolution with neutrino-emission modeling, including jet--environment interactions and relativistic effects such as beaming and time delays. A key feature of the present study is the implementation of zone-dependent particle transport and emission, combined with relativistic line-of-sight imaging. The focus is on neutrino emission from blob interactions with ambient photon fields, with explicit statements of acceleration zones, diffusion regimes, magnetic-field prescription, and target photon fields. This approach enables the generation of synthetic neutrino spectra and spatial intensity maps, which can be compared with the sensitivities of current and future neutrino observatories.

Finally, we provide an approximate assessment of detectability by relating the model predictions to published instrument sensitivities, a comparison that is simplified and sensitivity-limited.

This paper is organized as follows: Section~\ref{background} summarizes the acceleration and radiation assumptions. Section~\ref{neutrino_emissivity} gives the neutrino-emission formalism and target photon fields. Section~\ref{software} describes the software tools. Section~\ref{model_setup} presents the numerical setup and parameters. Section~\ref{results} gives the synthetic neutrino results and detectability estimates, and Section~\ref{conclusions} summarizes the main findings.

\section{Theoretical Background}
\label{background}

\subsection{Particle Acceleration Regions, Mechanisms, and Diffusion Regimes}
\label{sec:accel}

\noindent Terminology:
The overdense structures produced by intermittent injection and internal shocks are referred to here as blobs (or overdense internal structures). The term ``plasmoid'' is commonly used in plasma physics for magnetic islands produced by reconnection; when it appears in legacy figure filenames, it should be read in the present ``blob'' sense.

High-energy protons are accelerated at shock-like structures that naturally arise in intermittent relativistic jets. In the present work, we adopt a \emph{zone-based} description to explicitly specify where acceleration is assumed to occur and what transport regime is~used.

\noindent Zone A: Internal shocks/blob fronts:
Internal shocks form due to velocity irregularities and intermittency; in addition, blob fronts and compressed regions appear due to jet instabilities and jet--wind interactions \cite{Reynoso2008,Rieger2019}. In the RMHD output, candidate acceleration cells are identified by compression and shock proxies (e.g., negative velocity divergence \(\nabla\cdot \mathbf{v}<0\) together with elevated pressure gradients). These regions typically dominate the time-dependent nonthermal power injection.

\noindent Zone B: Recollimation/interaction layers:
As the jet propagates through ambient stellar and disk winds, recollimation and shear layers can form. These regions can host additional shocks and turbulence, contributing to acceleration and re-acceleration.

\noindent Zone C: Terminal/matter-loaded regions (head/cocoon):
Toward terminal regions of the jet, where matter may accumulate, thermal densities can become high. In such environments, proton--proton ($pp$) interactions may be relevant in general; however, they are not included in the emissivity calculation presented here. Transport in these regions may differ from that in internal blob fronts because of evolving turbulence levels and larger coherence scales.

\noindent Acceleration mechanism:
We assume first-order Fermi shock acceleration in the above shock-like regions. The acceleration timescale is modeled as follows:
\begin{equation}
t_{\mathrm{acc}}^{-1} = \eta \frac{c e B}{E_{p}},
\end{equation}
where \(\eta\) is an efficiency parameter (order \(0.01\)--\(0.1\) for relativistic shocks), \(B\) is the local magnetic field, and \(E_p\) the proton energy.

\noindent Diffusion regime and region-dependent transport:
Particle transport is controlled by the turbulence level and magnetic-field structure. At each particle energy \(E\) 
, we adopt an explicit diffusion coefficient \(D\) parameterization, with a reference value of \(D_0\) and reference energy \(E_0\):
\begin{equation}
D(E) = D_{0}\left(\frac{E}{E_{0}}\right)^{\delta},
\end{equation}
with \(\delta\) chosen \emph{by zone} to reflect different turbulence regimes. In strongly turbulent, shock-compressed blob fronts (Zone A), we use near-Bohm scaling (\(\delta \approx 1\)), whereas in more quiescent downstream / cocoon regions (Zones B/C), we adopt a weaker energy dependence (\(\delta \approx 1/2\) or \(\delta \approx 1/3\)). This zone dependence is motivated by recent discussions of microquasar-remnant transport and region-dependent diffusion \cite{Abaroa2025}. The adopted coefficients are used consistently in the emissivity pipeline and clearly stated in the model~setup.

\subsection{Magnetic Field Prescription in Acceleration Regions}
\label{sec:Bfield}

The magnetic field entering the acceleration and interaction rates is taken directly from the RMHD simulation output on a per-cell basis. To make the prescription explicit in acceleration zones, we use the following scheme:

\begin{itemize}
\item Primary: RMHD cell field: the local \(B\)-vector and magnitude are taken from the PLUTO RMHD solution in each cell.
\item Equipartition/turbulent amplification control (acceleration cells): in cells tagged as acceleration sites (Zones A/B), we may enforce or verify that the magnetic energy density satisfies
\begin{equation}
U_{B}=\frac{B^2}{8\pi} = \epsilon_{B}\,U_{\rm th},
\end{equation}
with \(\epsilon_B\) a model parameter (typically \(0.01\)--\(1\)). If the RMHD cell field is below the chosen equipartition fraction in a flagged acceleration cell, we adopt a conservative floor \(B \rightarrow \max(B_{\rm RMHD},B_{\rm eq})\), with \(B_{\rm eq}=\sqrt{8\pi \epsilon_B U_{\rm th}}\), U$_{th}$ being the local thermal energy density. This makes the assumed field strength in acceleration regions explicit and~\mbox{reproducible}.
\end{itemize}

At the jet base, the initial toroidal field is \(B_0 = 10^4~\mathrm{G}\), and the field evolves self-consistently in the RMHD simulation. The above prescription clarifies how \(B\) is interpreted/used in acceleration and emission calculations.

\subsection{Nonthermal Proton Distribution}
\label{sec:protons}

Nonthermal protons are described by a power-law energy distribution,
\begin{equation}
N_{p}(E) = K\,n_{p,0}\, E^{-\alpha},
\end{equation}
with spectral index \(\alpha \approx 2\) \cite{Reynoso2008}. The normalization factor \(K \ll 1\) specifies the fraction of hot protons relative to the thermal background density \(n_{p,0}\). Proton energies extend up to \(E_{\max} = 10^{6}\) GeV.

\medskip
\noindent Justification of \(E_{\max}\):
The adopted cutoff \(E_{\max}=10^{6}\,\mathrm{GeV}\) is motivated by limiting acceleration in the inner jet emission region resolved here, where maximum energies are constrained by finite residence time and energy losses. A Hillas-like estimate gives \(E_{\max}\simeq \eta eBR\beta\), where \(\eta<1\) is the acceleration efficiency, \(e\) is the proton charge, \(\beta\) is the local velocity in units of the speed of light, \(B\) is the local magnetic field, and \(R\) is the characteristic size of the acceleration region, taken here to be the blob scale in the inner jet. In the present framework, \(E_{\max}\) is treated as the physically motivated cutoff entering the emissivity calculation (Sections~\ref{neutrino_emissivity}--\ref{model_setup}) rather than as an unbounded cosmic-ray spectrum.

Protons are assumed isotropic in the jet comoving frame when the scattering length is shorter than the relevant loss lengths \cite{Rieger2019}. Energy losses such as synchrotron cooling and adiabatic expansion affect the steady-state proton spectrum \cite{Reynoso2008}.

\begin{figure}[H]
\includegraphics[width=10.5 cm]{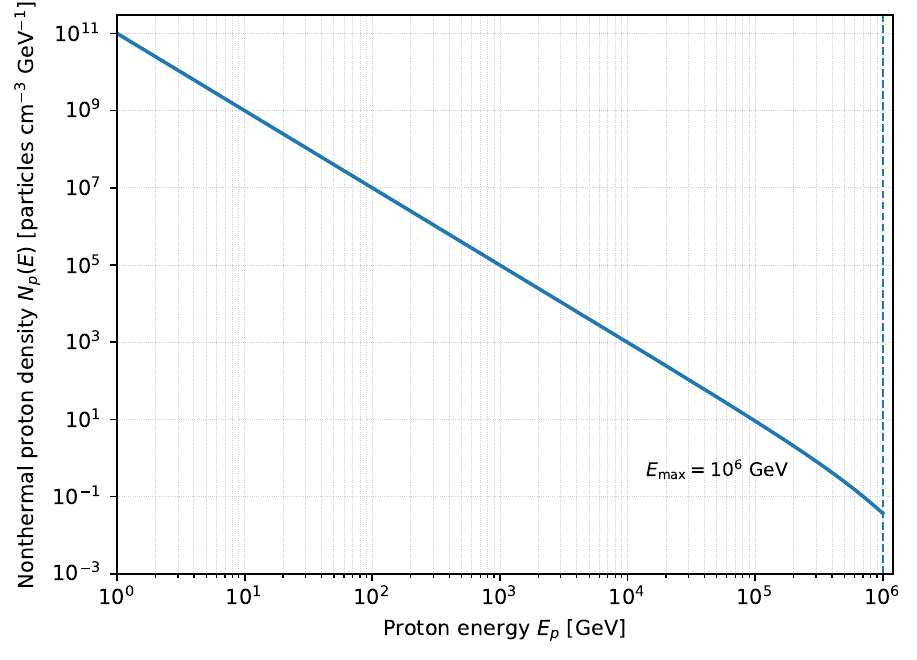}
\caption{Power-law 
 density of nonthermal protons used as input to the proton--photon emissivity calculation. The power-law shape is motivated by the first-order Fermi shock-acceleration prescription adopted in Section~\ref{sec:accel}, which yields a nonthermal proton spectrum with spectral index close to \(\alpha\simeq2\) as used here \cite{Reynoso2008}. The vertical axis gives the differential nonthermal proton density in particles cm$^{-3}$ GeV$^{-1}$. The plotted energy range is the one used in the calculation; the blue dashed vertical line marks the adopted cutoff at \(E_{\max}=10^{6}\,\mathrm{GeV}\).}
\label{fast-p-density}
\end{figure}

Figure~\ref{fast-p-density} shows the nonthermal proton distribution used as input to the proton--photon emissivity calculation.

\subsection{Origin of Target Photon Fields}
\label{sec:photons}

To remove ambiguity regarding the origin of photon fields used in \(p\gamma\) interactions, we explicitly model the target photon field as the sum of physically motivated components:

\begin{enumerate}
\item Companion star field (dominant in many HMXBs):
We model the stellar radiation as a diluted blackbody with temperature \(T_\star\) and radius \(R_\star\). At a distance \(r\) from the star, the photon energy density scales as \(u_\star(r)\propto (R_\star/r)^2\). The photon number density can be written as follows:
\begin{equation}
n_\star(\epsilon,r)=\left(\frac{R_\star}{r}\right)^2 n_{\rm BB}(\epsilon,T_\star),
\end{equation}
where \(n_{\rm BB}\) is the blackbody photon number density per unit energy.

\item Accretion disk/corona component (when relevant):
We include an additional thermal (or quasi-thermal) disk component with effective temperature \(T_{\rm disk}\) and characteristic radius \(R_{\rm disk}\), and optionally a coronal power-law tail if needed for higher-energy targets. In practice, this component can be switched on or off depending on the system class and parameter choice; a simplified accretion-disk wind prescription is used where appropriate.

\item Scattered/wind photon field:
A fraction $f_{sc}$ of stellar/disk photons can be scattered in the wind environment, producing a more isotropized target field. We may treat this as a scaled component: \(n_{\rm sc}(\epsilon)=f_{\rm sc}\,n_\star(\epsilon,r)\), with \(f_{\rm sc}\ll 1\).
\end{enumerate}

For blob--wind collisions, the stationary lab-frame photon distribution \(n^{\ast}(\epsilon^{\ast})\) is Doppler-transformed into the jet comoving frame as described in Section~\ref{neutrino_emissivity}. Starred quantities denote lab-frame quantities in this notation.

\section{Neutrino Emissivity}
\label{neutrino_emissivity}

\subsection{Neutrinos Produced Within the Jet}
Neutrino production in the present calculation proceeds through pion production in proton--photon ($p\gamma$) interactions, followed by pion decay. Proton--proton ($pp$) interactions may operate in dense regions of microquasar environments, but they are not part of the modeled emissivity here. The proton distribution in each computational cell is Lorentz-transformed into the observer frame \cite{TR11}, and steady-state transport equations govern the pion and neutrino spectra \cite{Kelner2006,Lipari2007,Reynoso2008,Reynoso2009}.

\medskip 
\noindent Muon treatment:
The emissivity includes the standard charged-pion decay chain \(\pi^\pm\rightarrow \mu^\pm+\nu_\mu(\bar\nu_\mu)\) and the subsequent muon decay contribution to the neutrino yield as implemented in the semi-analytic prescriptions adopted from \cite{Kelner2006,Lipari2007,Reynoso2009}. A separate muon cooling transport equation is not solved in the present version.

The neutrino emissivity at neutrino energy \(E\) is given by:
\begin{equation}
Q_{\pi \rightarrow \nu}(E) = \int_{E}^{E_{\max}} dE_{\pi}\, t_{\pi}^{-1}(E_{\pi}) N_{\pi}(E_{\pi})
\frac{\Theta(1 - r_{\pi} - x)}{E_{\pi}(1 - r_{\pi})},
\end{equation}
where \(N_{\pi}\) is the steady-state pion density, \(t_{\pi}\) the pion decay timescale, \(x=E/E_{\pi}\), \(r_{\pi} = (m_{\mu}/m_{\pi})^{2}\), and \(\Theta\) the Heaviside function \cite{Reynoso2009,SK15}.

\subsection{Neutrino Emissivity from Blob--Wind Collision}
To include neutrino emission enhancement from blob collisions with ambient photon fields, the stationary photon distribution \(n^{\ast}(\epsilon^{\ast})\) in the lab frame is Doppler-transformed into the jet comoving frame:
\begin{equation}
n(\epsilon, \Omega) = \frac{n^{\ast}(\epsilon^{\ast})}{4 \pi \delta_{\ast}^{2}},
\end{equation}
where \(\delta_{\ast}\) is the Doppler factor calculated per cell using local velocity vector and line-of-sight angles, with the velocity reversed to represent incoming photons as seen in the jet frame~\cite{Botcher23}. The transformed photon distribution replaces the synchrotron photon field used in earlier approaches, and the target field is explicitly decomposed as in Section~\ref{sec:photons}.

\subsection{Target Photon Distribution and Doppler-Factor Cases Used in the Model}
\label{sec:photon_cases}

For the $p\gamma$ channel used in our SED calculations, we adopt a power-law target photon distribution in the lab/host frame and implement three prescriptions for the interaction Doppler factor $D_{\rm ph}$ (Cases (a)--(c)
), as used in the model. The target photon density entering the interaction kernel is written as follows:
\begin{equation}
n_{\rm ph}(E_{\rm ph}) = n_0 \left(\frac{E_{\rm ph}}{E_0}\right)^{-\alpha_{\rm ph}} D_{\rm ph}^{-2}.
\end{equation}
where $\alpha_{\rm ph}$ is the photon spectral index, with a nominal value of 2.

\noindent Case (a): constant Doppler factor.
A single preset Doppler factor is assumed throughout the emitting blob:
\begin{equation}
n_{\rm ph}^{(a)} =
n_0 \left(\frac{E_{\rm ph}}{E_0}\right)^{-\alpha_{\rm ph}} D_{\rm ph,0}^{-2}.
\end{equation}

\noindent Case (b): local Doppler factor (cell-based).
The Doppler factor is computed locally for each cell using the cell velocity components and LOS angles:
\begin{equation}
n_{\rm ph}^{(b)} =
n_0 \left(\frac{E_{\rm ph}}{E_0}\right)^{-\alpha_{\rm ph}}
\left[D(-u_{xx},-u_{yy},-u_{zz},\phi_1,\phi_2)\right]^{-2}.
\end{equation}

\noindent Case (c): face-on collision approximation.
The LOS angles are set to zero (head-on \mbox{approximation}):
\begin{equation}
n_{\rm ph}^{(c)} =
n_0 \left(\frac{E_{\rm ph}}{E_0}\right)^{-\alpha_{\rm ph}}
\left[D(-u_{xx},-u_{yy},-u_{zz},0,0)\right]^{-2}.
\end{equation}

\noindent Case (d): benchmark subset of case (b).

This case uses the same local Doppler-factor prescription as case (b), but restricts the calculation to 60 selected cells. It is included only as a numerical benchmark.

In all cases, the velocity vector is reversed in the Doppler-factor evaluation to represent photons entering the jet comoving frame (Appendix
~\ref{details}).

\subsection{Proton--Photon Component Contribution}
\label{sec:pgamma}

The emission model computes neutrino spectra from proton--photon collisions between accelerated protons and ambient photon fields.

\subsection{Proton--Photon Channel}

For the proton--photon channel, the interaction rate in an isotropic photon field is given by \cite{Reynoso2009,Botcher23}:
\begin{equation}
\omega_{p\gamma}(E_p)= \frac{c}{2\gamma_p^2}
\int_{\epsilon_{\rm th}}^{2 \epsilon \gamma_p} d\epsilon_r\,
\sigma_{p\gamma}(\epsilon_r)\,\epsilon_r
\int_{\epsilon_{\rm th}/(2\gamma_p)}^{\infty} d\epsilon\,
\frac{n_{\rm ph}(\epsilon)}{\epsilon^2},
\end{equation}
where $\gamma_p=E_p/(m_pc^2)$, $\epsilon_r$ is the photon energy in the proton rest frame, and $n_{\rm ph}(\epsilon)$ is the target photon density, obtained from cases (a)--(c) above; case (d) is only the 60-cell benchmark subset of case (b).

The pion injection rate is then \cite{Reynoso2009}:
\begin{equation}
Q^{(p\gamma)}_{\pi}(E_{\pi})=  N_{\pi}(E_{p}) \int_{E_{\pi}}^{E_{p,\max}} dE_p\,
N_p(E_p)\,\omega_{p\gamma}(E_p)\,\delta\!\left(E_{\pi}-\kappa_{\pi}E_p\right),
\end{equation}
where $\kappa_{\pi}$ is the average inelasticity, set here equal to 0.2, mapping proton to pion energy.

Then, the pion number density is \cite{Reynoso2009}

\begin{equation}
N_{\pi}(E_{\pi}) = \frac{1}{\abs{b_{\pi}(E_{\pi})}} \int_{E_{\pi}}^{E_{max}} dE' Q(E') \times \exp{[- \tau (E_{\pi},E')]}
\end{equation}
where the pion optical depth is given by: 

\begin{equation}
\tau _{\pi}(E',E) = \int_{E'}^{E} \frac{dE'' t^{-1}_{\pi}(E)}{\abs{b_{\pi}(E'')}}
\end{equation}

The resulting neutrino emissivity is finally \cite{Reynoso2009}:
\begin{equation}
Q^{(p\gamma)}_{\nu}(E_{\nu})=\int_{E_{\nu}}^{E_{\pi,\max}} dE_{\pi}\,
\frac{t^{-1}_{\pi}(E_{\pi})\,N_{\pi}(E_{\pi})}{E_{\pi}(1-r_{\pi})}\,
\Theta\!\left(1-r_{\pi}-\frac{E_{\nu}}{E_{\pi}}\right).
\end{equation}

\subsection{Velocity and Direction Filtering}
To optimize computational costs, cells with velocity vectors aligned close to the observer's line-of-sight and speeds exceeding \(0.1c\) are prioritized for neutrino emission calculations. This filtering reduces the number of emitting cells, balancing fidelity and performance. Case (d), the 60-cell subset, is kept only as a technical benchmark of the calculation under limited sampling.

\section{Computer Programs Used}
\label{software}

\subsection{rlos: Relativistic Line of Sight Imaging}
The \texttt{rlos} code \cite{rlos}, developed by the author, performs special relativistic imaging by tracing rays through 4D RMHD simulation data, including relativistic beaming and time-delay effects.

\subsection{\texttt{PLUTO} Hydrocode} 
\texttt{PLUTO} \cite{Mignone2007} is a shock-capturing, finite-volume RMHD code used here for the jet simulations on structured 3D meshes.

\subsection{\texttt{nemiss}}
\texttt{nemiss} \cite{nemiss,Smponias_2021}, developed by the author, computes neutrino emissivities from hydrodynamic outputs, solving proton-to-neutrino cascades.

\subsection{Additional Tools}
Data visualization used \texttt{Veusz}. The codes are publicly available: \texttt{PLUTO} under \texttt{GPL}, \texttt{nemiss} and \texttt{rlos} under \texttt{LGPL}.

\section{Model Setup}
\label{model_setup}

Our RMHD simulations model intermittent relativistic twin microquasar jets at \(0.8c\). The jets propagate into ambient stellar and disk winds, with the companion star located outside the domain \cite{Fabrika2004}. The initial magnetic field is toroidal with \(10^{4}\, \mathrm{G}\) strength at the jet~base.

 \medskip
 \noindent Gravity and scales:
A twin-relativistic-jet binary stellar system, a microquasar, is simulated here. Jets emerge from the vicinity of the compact object, which orbits a main sequence star. The computational space focuses in the inner jet region, where neutrino emission is typically expected to occur. The companion star lies outside the computational box, and its stellar wind affects the system.

The simulations are special-relativistic RMHD (no gravitational potential is included). The adopted initial field \(B_0=10^4\,\mathrm{G}\) refers to the inner-jet injection region (cell size 10$^{10}$ cm) resolved in this homogeneous, for simplicity, RMHD domain (neutrino-emission-region scale), and not to radio-jet scales. B corresponds to the inner jet launching region and decreases rapidly with distance; radio-emitting scales are not modelled here.

 \medskip
 \noindent Initial distributions:
 At injection, the jet density is uniform within the nozzle (Table~\ref{Table-for-run-data}) and embedded in stratified ambient winds (stellar + disk wind). The magnetic field is initially toroidal within the injected jet and vanishes in the ambient medium. While a purely toroidal field is not strictly force-free in isolation, pressure gradients and the subsequent RMHD evolution provide the required force balance in the numerical setup. Ambient wind density is falling off as 1/r$^2$ away from the jet base. Accretion disk wind density falls off as 1/r away from the equatorial plane.

\noindent Explicit model choices:
The acceleration zones are defined as in Section~\ref{sec:accel}. We use a zone-dependent diffusion parameterization \(D(E)=D_0 (E/E_0)^\delta\), adopting near-Bohm scaling (\(\delta\simeq 1\)) in blob/shock regions and weaker scaling (\(\delta\simeq 1/2\) or \(1/3\)) in downstream/cocoon regions (see Section~\ref{sec:accel}). The magnetic field used in acceleration and interaction rates is taken from the RMHD cell field, with an optional equipartition floor in tagged acceleration cells (Section~\ref{sec:Bfield}). Target photon fields are explicitly decomposed (Section~\ref{sec:photons}) and transformed to the comoving frame where needed. For the $p\gamma$ SED cases, we adopt the Doppler prescriptions given in Section~\ref{sec:photon_cases}; case (d) is only a 60-cell benchmark subset of case (b).

The computational mesh is \(60 \times 100 \times 50\) Cartesian cells, each \(2 \times 10^{10}\) cm in length. The simulation parameters are summarized in Table~\ref{Table-for-run-data}. Neutrino line-of-sight imaging uses local velocity and magnetic field data per cell, with relativistic effects modeled.

\begin{table}[H]
\caption{Simulation and model parameters.}
\label{Table-for-run-data}
\begin{tabularx}{\textwidth}{LLL}
\toprule
\textbf{Parameter} & \textbf{Value} & \textbf{Comments} \\
\midrule
Cell size \(l_{\mathrm{cell}}\) & \(2.0 \times 10^{10}\) cm & PLUTO cell length \\
Jet density \(\rho_{\mathrm{jet}}\) & \(1.0 \times 10^{11} \mathrm{cm}^{-3}\) & Jet proton density \\
Wind max density \(\rho_{w}\) & \(1.0 \times 10^{13} \mathrm{cm}^{-3}\) & Max ambient wind density \\
Disk wind max density \(\rho_{dw}\) & \(2.0 \times 10^{13} \mathrm{cm}^{-3}\) & Max disk wind density \\
Time unit & 1 s & Model time scale \\
Max run time & 204 s & Total simulation time \\
Integrator & MUSCL-Hancock & PLUTO scheme \\
Physics & Special Relativistic MHD & RMHD setup \\
Initial magnetic field & \(10^{4}\) G & Toroidal field \\
Binary separation & \(4.0 \times 10^{12}\) cm & \\
Jet kinetic luminosity \(L_k\) & \(2 \times 10^{38}\) erg/s & Jet power \\
Jet speed \(\beta\) & 0.8 & Jet velocity fraction of \(c\) \\
Grid size & \(60 \times 100 \times 50\) & PLUTO mesh size \\
Imaging method & Focused beam & rlos configuration \\
Imaging plane & YZ plane & Fiducial screen orientation \\
Emission type & Neutrinos & Synthetic emission mode \\
\bottomrule
\end{tabularx}
\end{table}
\section{Results and Discussion}
\label{results}

\begin{figure}[H]
\includegraphics[width=10.5cm]{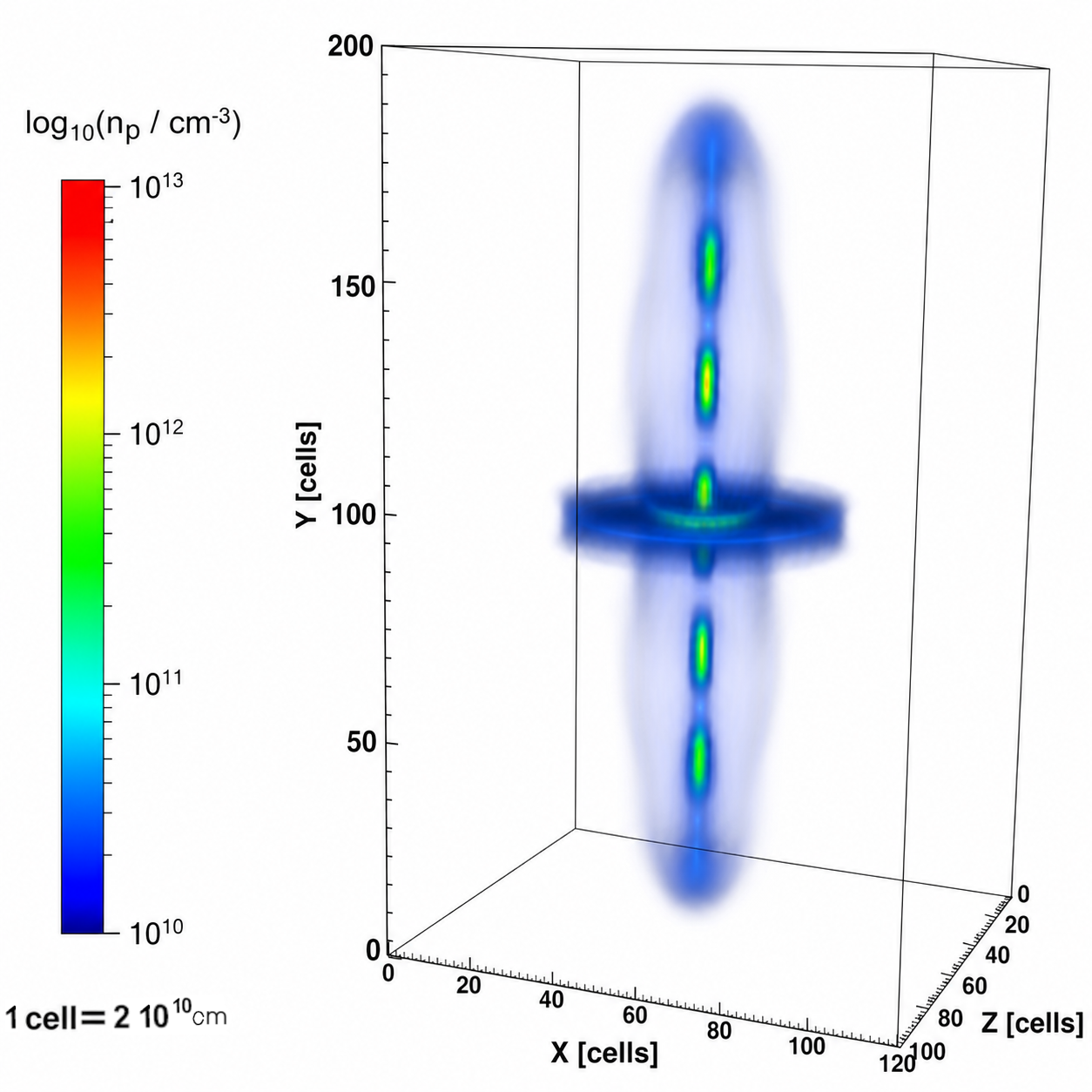}
\caption{Three-dimensional 
 volume rendering of the PLUTO RMHD bulk-proton number density n$_p$, at model time t = 50 s. The color scale is purely logarithmic and represents \(n_p\) in \(\mathrm{cm^{-3}}\). The axes are given in grid cell lengths, whereas 1 cell length = 2~$\times$~10$^{10} ~\mathrm{cm} $. Intermittent twin jet blobs traverse ambient stellar and disk winds, sweep aside surrounding matter, and produce dynamic equatorial structures. For the currently employed emission mechanism, which includes jet interactions with surrounding photon fields, regions of interest for emission include mainly the jets' heads, where jet matter strongly interacts with ambient photons. The jet heads are shown at a stage where they are nearly exiting the grid in order to better display the structure of the twin jets' system.}
\label{neutrinoscalejet}
\end{figure}

\begin{figure}[H]
\centering

\begin{adjustwidth}{-\extralength}{0cm}
\includegraphics[width=15.0cm]{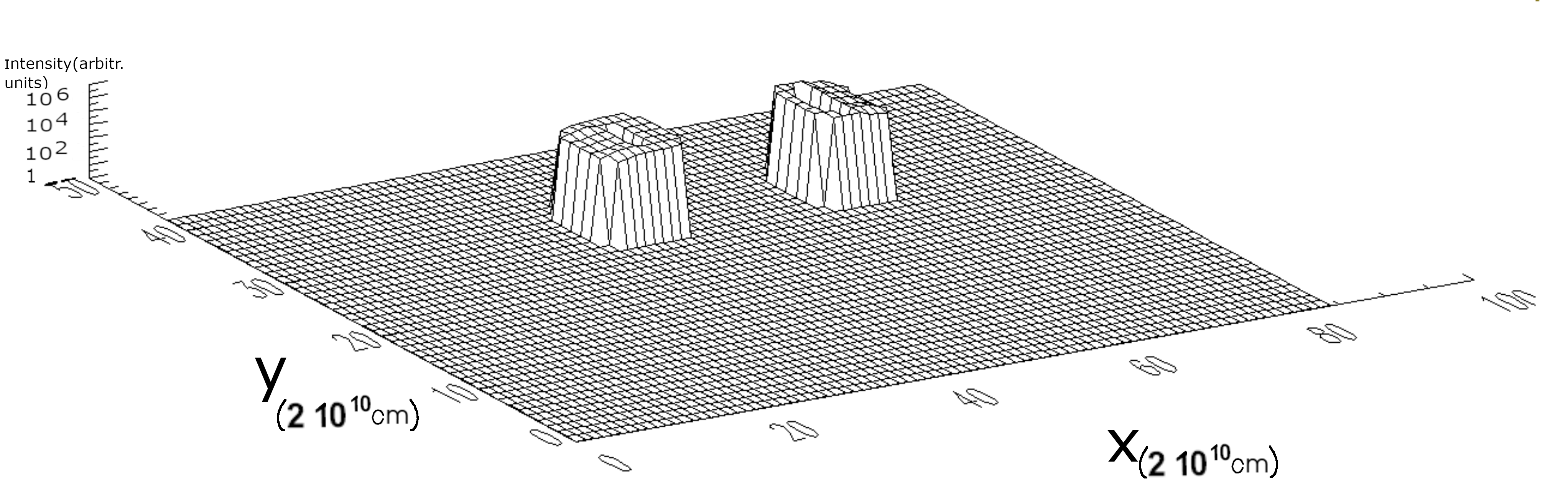}
\end{adjustwidth}
\caption{Synthetic 
 neutrino image at model time t = 30 s and \(E_\nu\simeq1.1\) GeV.
The X and Y axes are in computational-cell length units, with \(1\) cell length \(=2\times10^{10}\,\mathrm{cm}\).
The image is obtained by applying the  \texttt{rlos} ray-tracing code to a PLUTO RMHD snapshot while viewing the system from the side of the computational box. The y axis here corresponds to the x axis of the hydrodynamic simulation, while the x axis here corresponds to the y axis there. The shot used is taken somewhat earlier than the one shown in Figure~\ref{neutrinoscalejet} in order to better catch the initial dynamic and radiative effect caused by the injection. The two intensity maxima correspond to the two opposite leading jet blobs, at the jets' heads, visible in the density rendering. The representative energy \(E_\nu\simeq1.1\)~GeV is used only for morphology visualization, because the modeled low-energy emissivity makes the spatial structures easier to resolve. Intensity is shown in arbitrary logarithmic units and is used only as a morphology tracer.}
\label{neutrinobothplots}
\end{figure}

Intermittent twin jets propagate through ambient matter, generating dynamic equatorial structures as jet blobs interact with stellar and disk winds (Figure~\ref{neutrinoscalejet}). Emission from the model, as a result of the PLUTO--nemiss--rlos pipeline, is shown in Figure~\ref{neutrinobothplots}, for a sample~energy.


Figure~\ref{neutrinobothplots} is plotted at \(E_\nu\simeq1.1\)~GeV only to illustrate morphology. This energy is used because the modeled emissivity is higher at low energies, making the spatial emission features easier to resolve; it is not intended as a detector-sensitivity energy. Actual detectability is discussed only above TeV energies, where neutrino detectors have sensitivity.

\subsection{Proton--Photon Spectra and Detectability}
The results below refer to the proton--photon component only. Proton--proton interactions are a possible neutrino production process in dense microquasar environments, but they are not included in the present model. The aim here is to examine how the adopted target photon field and Doppler prescriptions affect the modeled proton--photon spectra and their comparison with detector sensitivities. The accompanying escaping gamma-ray cascade spectrum is not computed in this version. However, a production-level neutral-pion gamma-ray counterpart and an approximate neutrino-to-gamma-ray energy-flux ratio are estimated below.

\vspace{-14pt}

\subsection{Flux Normalization, Instrument Sensitivities, and Distance-Dependent Detectability}


A realistic 
 assessment of detectability requires accounting for atmospheric neutrino backgrounds, angular resolution, exposure, and analysis cuts. The sensitivity overlays and the derived distance horizon \(d_{\max}(E)\) presented below should therefore be interpreted as sensitivity-limited illustrative metrics rather than a full background-limited discovery~\mbox{potential}.

\begin{figure}[H]
\vspace{-4pt}
\includegraphics[width=10.5cm]{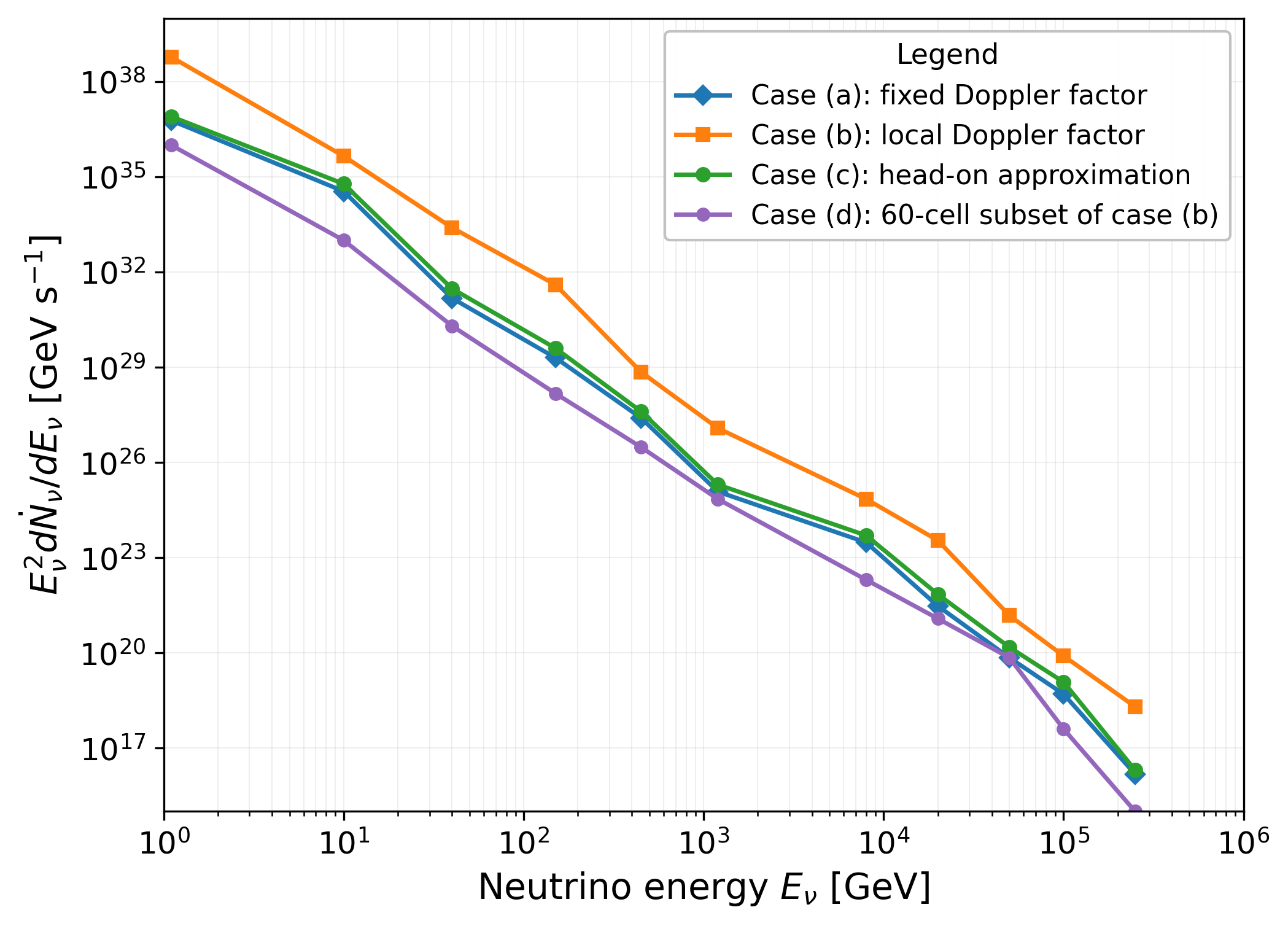}
\caption{Source-level neutrino spectral power for the proton--photon (\(p\gamma\)) component, shown as \(E_\nu^2 d\dot{N}_\nu/dE_\nu\) in \(\mathrm{GeV\,s^{-1}}\). Cases (a)--(c) correspond to the Doppler prescriptions adopted for the target photon field: fixed Doppler factor, local Doppler factor, and head-on approximation, respectively (Section~\ref{sec:photon_cases}). The normalization assumes a neutrino efficiency \(\eta_\nu=0.01\) relative to the jet kinetic power \(L_k=2\times10^{38}\,\mathrm{erg\,s^{-1}}\) listed in Table~\ref{Table-for-run-data}, giving \(L_\nu=2\times10^{36}\,\mathrm{erg\,s^{-1}}\). The conversion to GeV units uses \(1\,\mathrm{erg}=624\,\mathrm{GeV}\). Case (d) is a 60-cell benchmark subset of case (b) and is included only as a numerical sampling test.}
\label{sed_plot}
\end{figure}

\begin{figure}[H]
\includegraphics[width=12.5cm]{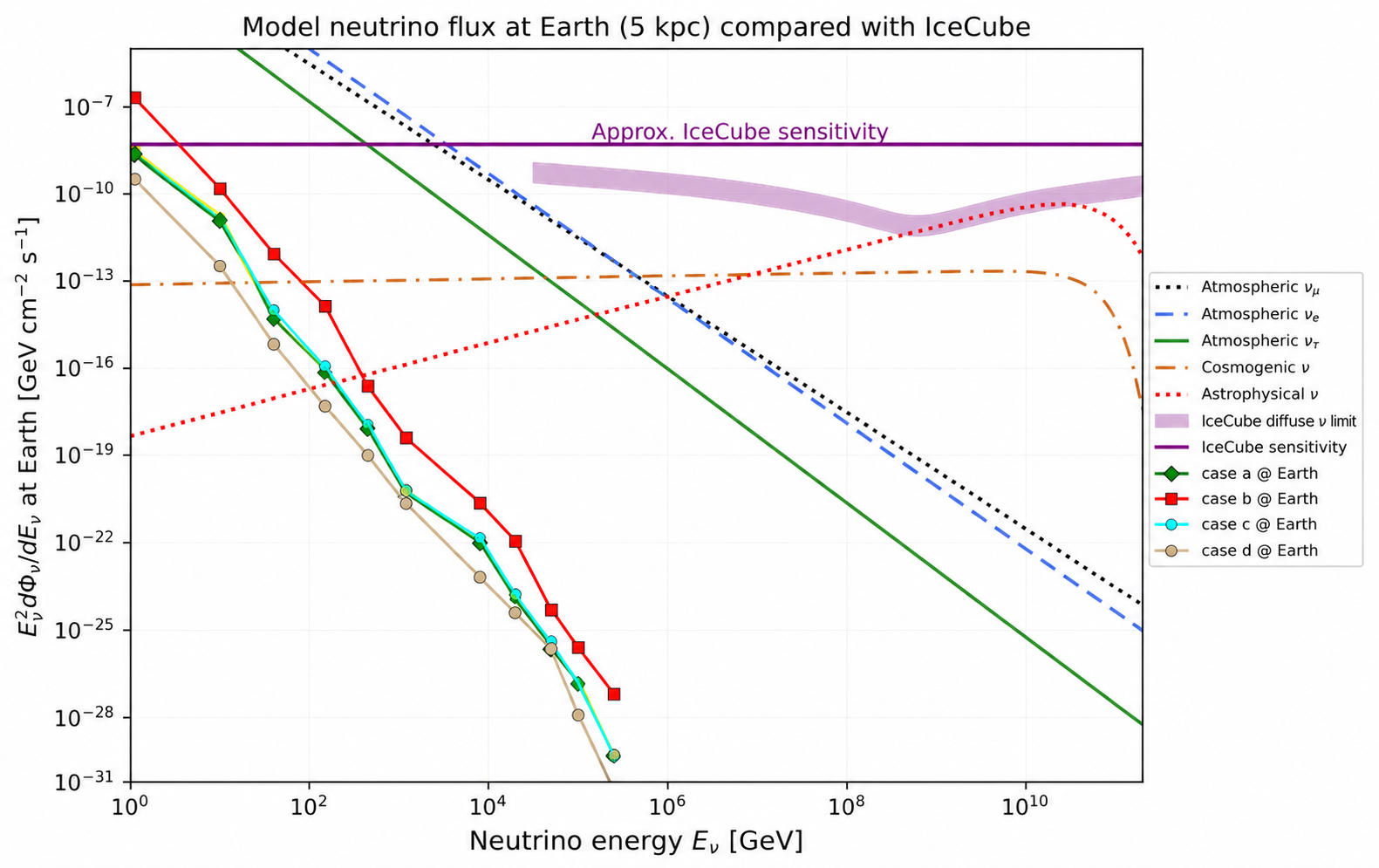}
\caption{Modeled 
 proton--photon neutrino component scaled to Earth at \(d=5\) kpc and compared with an approximate IceCube sensitivity reference. The source-level spectra of Figure~\ref{sed_plot} are divided by \(4\pi d^2\). In this normalization, the modeled component remains below the approximate IceCube sensitivity threshold over the detector-relevant energy range.}
\label{fig:norm}
\end{figure}

\begin{figure}[H]
\includegraphics[width=10.5cm]{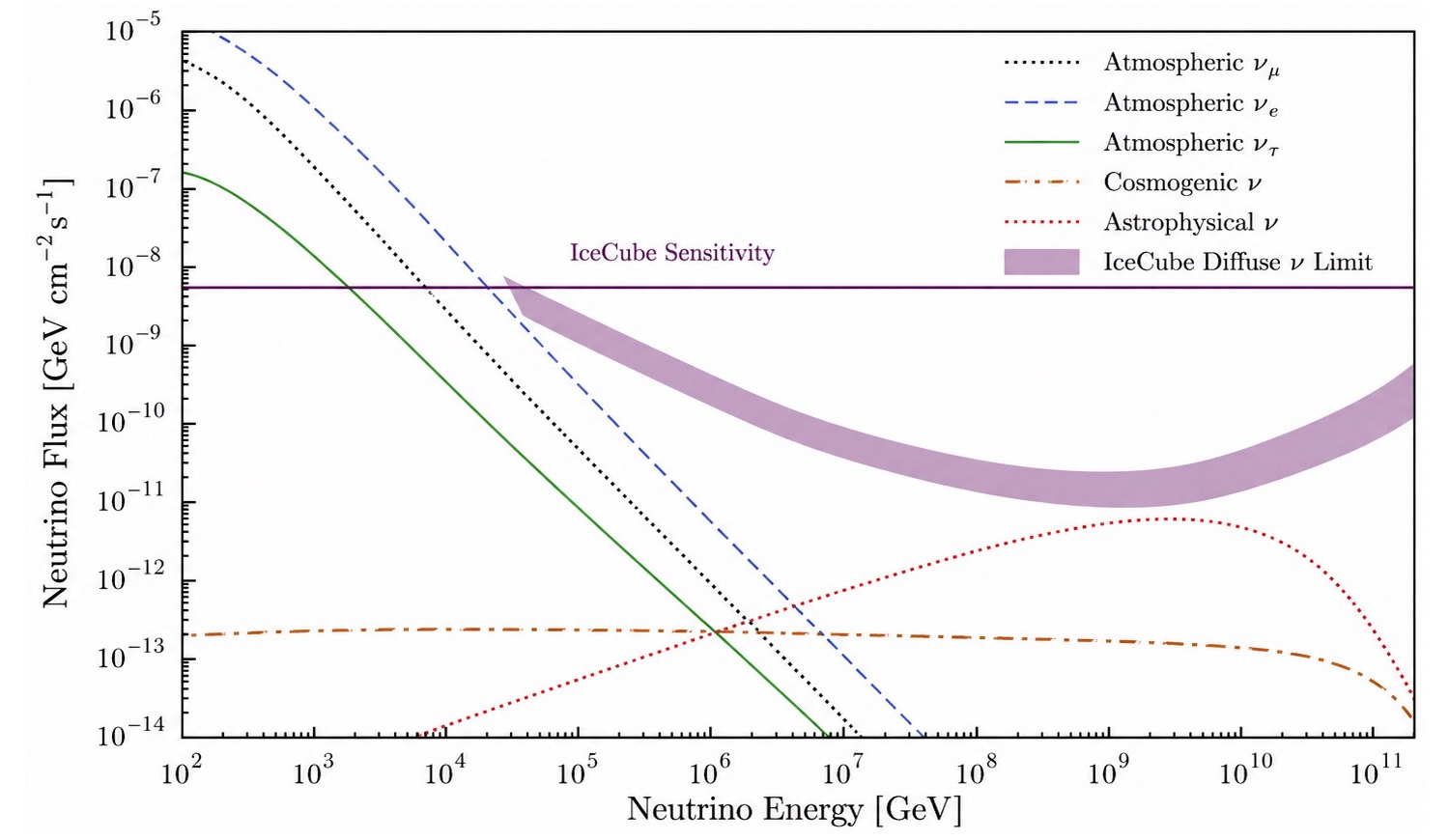}
\caption{Representative neutrino flux components as a function of neutrino energy. The atmospheric neutrino fluxes ($\nu_\mu$, $\nu_e$, $\nu_\tau$) correspond to production in cosmic-ray air showers and follow standard parameterizations \cite{Honda2007,Gaisser2016}. The cosmogenic neutrino flux arises from ultra-high-energy cosmic ray interactions with the cosmic microwave background (GZK mechanism) \cite{Beresinsky1969,Ahlers2010}. The astrophysical neutrino component represents the diffuse high-energy flux measured by IceCube \cite{Aartsen2015}. The shaded region indicates the IceCube diffuse neutrino flux limits, while the horizontal line provides an approximate indication of the detector sensitivity. The curves are schematic but reflect typical magnitudes and spectral behavior reported in the literature. The IceCube astrophysical component and the corresponding reference level are based on the published northern-sky muon-neutrino analysis of IceCube, using 659.5 days of live time recorded between May 2010 and May 2012. The horizontal IceCube sensitivity line is an approximate benchmark added for visual comparison; it is not a newly derived detector response curve and should not be interpreted as a discovery threshold for a specific exposure.}
\label{fig:distance_horizon}
\end{figure}

Figures~\ref{sed_plot} and~\ref{fig:norm} present the model-based neutrino emission results and their detectability comparison. Figure~\ref{sed_plot} shows the predicted proton--photon spectra for the adopted emission prescriptions, while Figure~\ref{fig:norm} compares the rescaled flux with an approximate IceCube sensitivity reference. Figure~\ref{fig:distance_horizon} provides a schematic context for representative neutrino flux components and IceCube sensitivity.

The curves in Figure~\ref{fig:distance_horizon} are not fitted in the present work. They are literature-based reference components plotted for context. The atmospheric components are represented by standard steep power-law parameterizations:
\begin{equation}
\Phi_{\rm atm}^{i}(E_\nu)=A_i\left(\frac{E_\nu}{E_0}\right)^{-\gamma_i},
\end{equation}
Here, 
 \(i\) labels the neutrino flavour \(\nu_\mu,\nu_e,\nu_\tau\), \(E_0\) is a reference energy, \(A_i\) is the normalization for each flavour, and \(\gamma_i\) is the corresponding spectral index. The normalizations and slopes are chosen to reproduce the published atmospheric-neutrino curves of  \cite{Honda2007,Gaisser2016}. The astrophysical IceCube component is represented by a single-flavour power law:
\begin{equation}
\Phi_{\rm astro}(E_\nu)=\Phi_0
\left(\frac{E_\nu}{100\,{\rm TeV}}\right)^{-\Gamma},
\end{equation}
Here, \(\Phi_0\) is the single-flavour normalization and \(\Gamma\) is the astrophysical spectral index. The expression uses the published IceCube normalization and spectral index. The cosmogenic component follows the GZK template curves cited in the figure caption.

More specifically, for the IceCube astrophysical component, we use the published form
\begin{equation}
\Phi(E_\nu)=9.9\times10^{-19}
\left(\frac{E_\nu}{100\,{\rm TeV}}\right)^{-2}
{\rm GeV^{-1}\,cm^{-2}\,s^{-1}\,sr^{-1}}
\end{equation}
per flavour, as reported for the 659.5-day northern-sky sample.

We emphasize that the sensitivity curves shown correspond to published detector performance estimates, which already account for atmospheric neutrino backgrounds under standard analysis assumptions. Therefore, the comparison presented here should be interpreted as a sensitivity-limited detectability estimate rather than a full likelihood-based background analysis. A dedicated background-limited discovery analysis would require detector-specific Monte Carlo simulations and event-selection modeling, which are beyond the scope of the present source-modeling study.


To assess detectability, the observable neutrino flux is obtained from the model luminosity via an assumed source distance \(d\). Fluxes scale as follows:
\begin{equation}
\Phi_\nu(E;d)=\Phi_\nu(E;d_0)\left(\frac{d_0}{d}\right)^2.
\end{equation}
In figures that include observables, we overlay sensitivity curves for relevant instruments (e.g., IceCube) and we explicitly state the assumed distance used for flux conversion (here, we use a canonical galactic distance \(d_0=5~\mathrm{kpc}\) unless otherwise indicated).

A convenient distance-horizon metric at each energy is:
\begin{equation}
d_{\max}(E) = d_0 \sqrt{\frac{\Phi_\nu(E;d_0)}{\Phi_{\rm sens}(E)}},
\end{equation}
where \(\Phi_{\rm sens}(E)\) is the instrument sensitivity at energy \(E\). This provides a detectability-versus-distance statement independent of a single assumed distance.


\vspace{-10pt}

Figure~\ref{fig:norm} compares the modeled proton--photon component with an approximate IceCube sensitivity reference after scaling the source emission to Earth for the assumed distance. The flux is divided by \(4\pi d^2\), as described in Appendix~\ref{app:norm}. In this normalization, the modeled proton--photon component lies below the approximate IceCube sensitivity threshold across the detector-relevant energy range. Much of the model power also appears at energies below the main IceCube sensitivity window. Relativistic beaming may enhance the apparent flux for favorable viewing geometries, but no geometric enhancement sufficient to close the full gap is claimed here. A detector-specific discovery analysis is beyond the scope of the present work.

The fluxes shown in Figures~\ref{sed_plot} and~\ref{fig:norm} should be interpreted as a fiducial microquasar calculation rather than as a source-specific prediction. They scale approximately as follows:
\begin{equation}
E_\nu^2\Phi_\nu(E_\nu;d)
\propto
\frac{\eta_\nu L_k}{4\pi d^2},
\end{equation}
Here, \(E_\nu^2\Phi_\nu\) denotes the spectral energy flux at distance \(d\), \(\eta_\nu\) is the assumed neutrino efficiency, and \(L_k\) is the jet kinetic power. The scaling also has additional dependence on the target photon density, Doppler factor, viewing angle, maximum proton energy, and acceleration-zone filling factor. Therefore, differences between the present curves and other modeled galactic microquasar fluxes can arise from different assumed distances, jet kinetic powers, photon-field densities, Doppler prescriptions, and whether the dominant channel is \(p\gamma\) or \(pp\). The present calculation isolates the \(p\gamma\) contribution; models including dense-target \(pp\) interactions may produce different flux levels and spectral shapes.

\subsection{Comparison with Previous Galactic Microquasar Neutrino-Flux Models}

\subsubsection{General Comments}

The fluxes obtained in the present work should be interpreted as a fiducial proton--photon calculation rather than as a source-specific prediction. Previous models of galactic microquasar neutrino emission have often considered different physical regimes. For example, \cite{Christiansen2006} modeled high-energy neutrino production in X-ray binaries with relativistic jets interacting with dense stellar winds, where neutrinos and correlated gamma rays arise mainly from \(pp\) interactions. In that type of dense-wind scenario, the expected neutrino output can be larger because the target matter density is high.

Similarly, \cite{Reynoso2008} studied gamma-ray and neutrino production in the dark jets of SS433 through \(pp\) interactions, emphasizing the role of absorption and the gamma-ray-to-neutrino connection. The authors of \cite{Reynoso2009}  also showed that neutrino production close to the compact object can be strongly affected by synchrotron losses of secondary pions and muons, while interactions farther out in dense wind clumps may be more favorable for detectability.

The present calculation differs from those works because it isolates the \(p\gamma\) channel in transient inner-jet blobs interacting with ambient photon fields. The model flux therefore depends primarily on the accelerated proton normalization, target photon density, Doppler prescription, viewing geometry, maximum proton energy, and source distance. In contrast, \(pp\)-dominated models scale mainly with the density of matter targets. Consequently, differences between the present fluxes and previously modeled galactic microquasar fluxes are expected and mainly reflect the different interaction channel, target density, source geometry, and assumed jet power. The present result is therefore consistent with the broader picture that detectable microquasar neutrino emission is more favorable in dense \(pp\) environments, whereas the isolated \(p\gamma\) component considered here remains challenging to detect for the adopted parameters.

\subsubsection{Quantitative Comparison}

To make the requested comparison with previous galactic microquasar models explicit, Table~\ref{mq_flux_comparison} compares the present Case~(b) flux at Earth with representative published results. The values for the present work are read from Figure~\ref{fig:norm}. Literature values are quoted in the form provided by the original papers, or estimated from published plots where possible; consequently, they should be interpreted as order-of-magnitude comparisons rather than as a uniform re-analysis.

\begin{table}[H]
\caption{Order-of-magnitude 
 comparison of the present proton--photon flux with previous galactic microquasar neutrino-flux models. The comparison uses the spectral energy flux \(E_\nu^2\Phi_\nu\) in \(\mathrm{GeV\,cm^{-2}\,s^{-1}}\), where such a differential quantity is available. Values marked as approximate are read from published curves or converted from published integrated luminosities/event-rate estimates, and are therefore not detector-response-corrected in a uniform way.}
\label{mq_flux_comparison}
\begin{adjustwidth}{-\extralength}{0cm}
\centering
\footnotesize
\begin{tabularx}{\fulllength}{m{3cm}<{\raggedright}LLLLm{4.48cm}<{\raggedright}}
\toprule
\textbf{Model/Scenario} & \textbf{Main Channel} & \boldmath{\textbf{\(E_\nu^2\Phi_\nu(1\,\mathrm{TeV})\)}} & \boldmath{\textbf{\(E_\nu^2\Phi_\nu(10\,\mathrm{TeV})\)}} & \boldmath{\textbf{\(E_\nu^2\Phi_\nu(100\,\mathrm{TeV})\)}} & \textbf{Comment} \\
\midrule
This work, Case~(b), \(d=5\) kpc & \(p\gamma\) & \(\sim4\times10^{-19}\) & \(\sim1\times10^{-21}\) & \(\sim2\times10^{-26}\) & Isolated inner-jet photon-target component, read from Figure~\ref{fig:norm}. \\
\midrule
Christiansen et al.~\cite{Christiansen2006}; LS I~+61~303 dense wind & \(pp\) & \multicolumn{3}{p{7.0cm}}{\vspace{-15pt}Integrated \(>1\) TeV \(\nu_\mu\) luminosity \(\sim5\times10^{34}\,\mathrm{erg\,s^{-1}}\), corresponding to   a log-averaged energy flux of order \(10^{-8}\,\mathrm{GeV\,cm^{-2}\,s^{-1}}\) for \(d\simeq2\) kpc.} & Differential values at fixed energies are not tabulated; the model gives an IceCube-scale event-rate estimate rather than a directly comparable \(E^2\Phi\) triplet. \\

\bottomrule
\end{tabularx}
\end{adjustwidth}
\end{table}

\begin{table}[H]\ContinuedFloat
\caption{\textit{Cont.}}
\begin{adjustwidth}{-\extralength}{0cm}
\centering
\footnotesize
\begin{tabularx}{\fulllength}{m{3cm}<{\raggedright}LLLLm{4.48cm}<{\raggedright}}
\toprule
\textbf{Model/Scenario} & \textbf{Main Channel} & \boldmath{\textbf{\(E_\nu^2\Phi_\nu(1\,\mathrm{TeV})\)}} & \boldmath{\textbf{\(E_\nu^2\Phi_\nu(10\,\mathrm{TeV})\)}} & \boldmath{\textbf{\(E_\nu^2\Phi_\nu(100\,\mathrm{TeV})\)}} & \textbf{Comment} \\
\toprule

Reynoso et al.~\cite{Reynoso2008};\linebreak   SS433 dark jets & \(pp\) & \(\sim10^{-8}\) & \(\sim10^{-9}\) & \(\sim10^{-12}\) & Approximate values read from their phase-dependent differential-flux map for \(q_{\rm rel}=10^{-4}\); the normalization is source-specific and tied to dense baryonic jets. \\

\midrule
Reynoso and Romero~\cite{Reynoso2009}; magnetized jet/wind-clump cases & \(p\gamma\), \(pp\) & model-dependent & model-dependent & model-dependent & Their published fluxes depend strongly on secondary pion/muon losses and on whether the interaction occurs near the compact object or in dense wind clumps; no single-source-independent flux triplet is~quoted. \\
\bottomrule
\end{tabularx}
\end{adjustwidth}
\end{table}

The comparison shows why the present flux is lower than many dense-wind or dark-jet estimates. Those models often assume efficient \(pp\) interactions in dense matter targets, whereas the present calculation isolates the \(p\gamma\) contribution from transient inner-jet blobs interacting with ambient photon fields. The main physical source of the difference is therefore the target density and interaction channel. As mentioned above, additional dependence lies on source distance, jet power, Doppler prescription, viewing angle, and acceleration-zone filling factor.

Our detectability estimate is intended as an illustrative, sensitivity-limited estimate based on published instrument performance curves. Since different experiments adopt different significance criteria, background treatments, exposure assumptions, and energy binning schemes, the curves should not be interpreted as strictly uniform discovery~\mbox{thresholds}. 

\subsection{Associated $\gamma$-Ray Spectrum}

\begin{figure}[H]
\includegraphics[width=12.5cm]{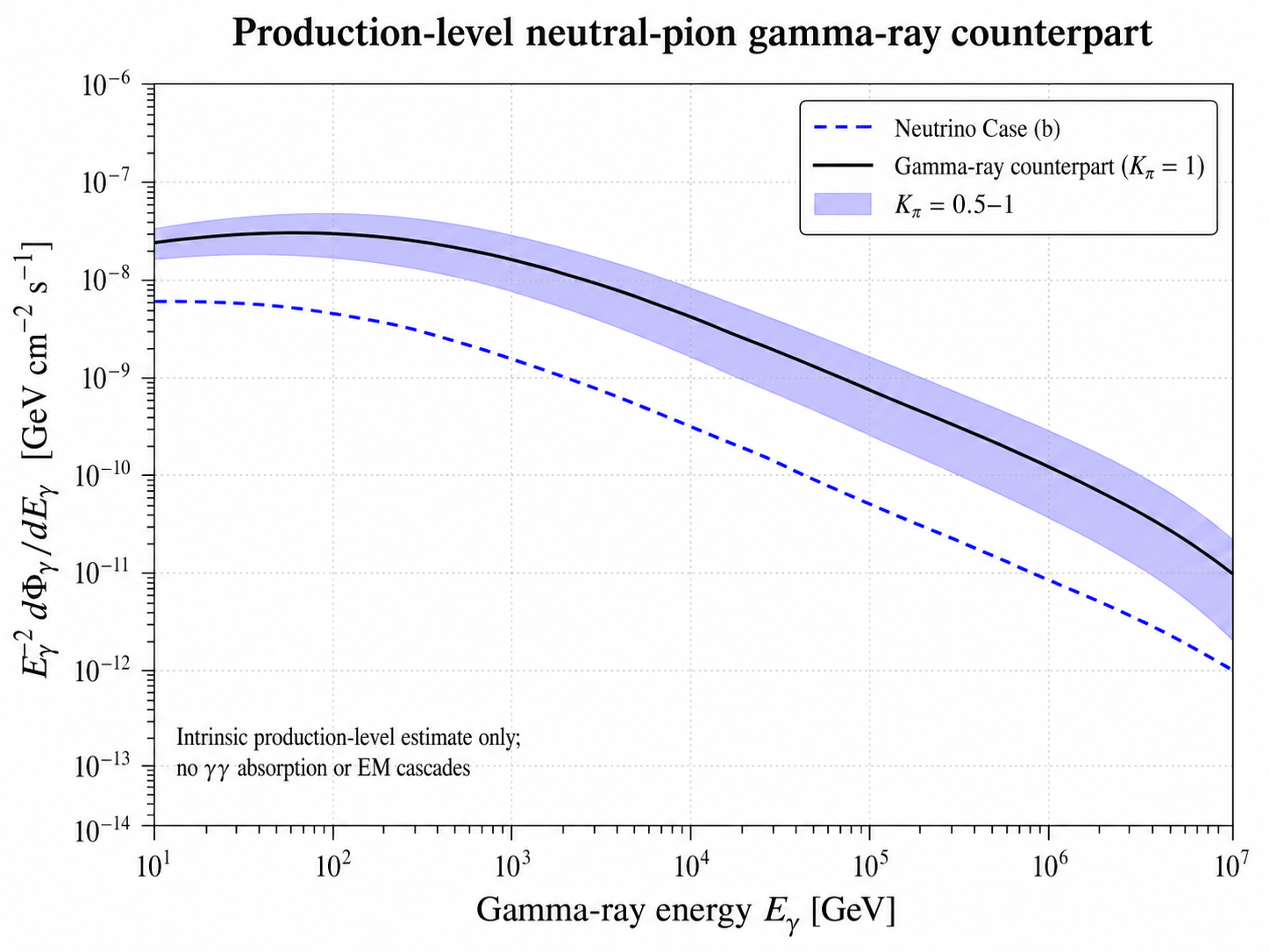}
\caption{Production-level neutral-pion gamma-ray counterpart inferred from the proton--photon Case~(b) neutrino spectrum. The conversion uses \(E_\gamma\simeq2E_\nu\) and \(E_\gamma^2\Phi_\gamma\simeq 4E_\nu^2\Phi_{\nu,\mathrm{all}}/(3K_\pi)\), with \(K_\pi=0.5-1\). The shaded band shows the uncertainty associated with this charged-to-neutral pion ratio. This is an intrinsic production-level estimate only; internal \(\gamma\gamma\) absorption and electromagnetic cascading are not included; so, the curve should not be interpreted as the escaping GeV--TeV gamma-ray spectrum. It is not compared directly to observed microquasar spectra.}
\label{fig:gamma}
\end{figure}

The present calculation follows the neutrino output of the photohadronic channel and does not compute a full gamma-ray radiative-transfer spectrum. Nevertheless, the associated gamma-ray component can be estimated at the production level. In $p\gamma$ interactions, charged pion production, which leads to neutrinos, is accompanied by neutral pion production, which leads to gamma rays. For pion-decay emission, one may then write, approximately:
\[
E_\nu^2 \Phi_{\nu,\mathrm{all}}(E_\nu)
\simeq
\frac{3K_\pi}{4}
E_\gamma^2 \Phi_\gamma(E_\gamma),
\qquad E_\gamma \simeq 2E_\nu ,
\]
where \(K_\pi=N_{\pi^\pm}/N_{\pi^0}\) is the charged-to-neutral pion ratio. For photohadronic interactions, \(K_\pi\sim0.5-1\), giving a production-level all-flavour neutrino-to-gamma-ray energy-flux ratio of order \(0.4-0.8\). The corresponding single-flavour ratio is smaller by about a factor of three. Therefore, the intrinsic neutral-pion gamma-ray component would be expected to be of the same order as the neutrino component. Figure~\ref{fig:gamma} shows the corresponding production-level gamma-ray counterpart for case~(b).

However, this estimate should not be interpreted as the observable gamma-ray spectrum. High-energy photons may undergo internal \(\gamma\gamma\) absorption and electromagnetic cascading before escaping the source. A quantitative comparison with observed GeV--TeV gamma-ray spectra of microquasars would therefore require additional photon transport and cascade modeling, which is left for future work.

\section{Conclusions}
\label{conclusions}

We present a time-resolved model for proton--photon neutrino emission from relativistic microquasar jets, combining RMHD simulations with explicit prescriptions for acceleration zones, zone-dependent diffusion, magnetic-field treatment, and target photon fields. Proton--proton interactions are relevant in dense environments and are mentioned for context, but they are not modeled here. 

In more dense environments, neutrino emission by proton--proton interactions in the jet is possible, but the density might be too large to allow for gamma-ray emission from the pion decays. On the other hand, in less-dense regions, proton--photon interactions could produce neutrinos, while the overall opacity might be low enough to allow detectable gamma-rays to escape the system.

In our case, for the adopted normalization and a distance of 5 kpc, the calculated proton--photon component lies below the approximate IceCube sensitivity reference, especially in the detector-relevant energy range. Favorable viewing geometry and relativistic beaming may increase the apparent flux, but the present comparison should be regarded as a sensitivity-limited estimate rather than a discovery forecast.  

Although a neutral-pion gamma-ray counterpart is expected at production, its escaping GeV--TeV spectrum cannot be inferred without modeling internal \(\gamma\gamma\) absorption and electromagnetic cascading, since high-energy photons may be absorbed and cascaded inside the source. Under the assumptions adopted here, the present results therefore indicate that detectable neutrino emission, from the modeled \(p\gamma\) channel in microquasars, is challenging, and is likely not going to be a viable mechanism for generating simultaneous neutrino-gamma emission.

Future work may explore higher resolution simulations, refined radiation-field geometries, and more detailed system-specific parameter sets. 
 
\vspace{6pt}


\funding{This research received no external funding.}
\institutionalreview{Not applicable.}
\informedconsent{Not applicable.}
\dataavailability{The original contributions presented in this study are included in the article. Further inquiries can be directed to the corresponding author. 
}
\acknowledgments{The author thanks colleagues for valuable comments on this manuscript. Special thanks go to Gustavo Romero 
 for his insightful comments on an earlier version of this work. The author also gratefully acknowledges important comments from Ralph Spencer.}

\conflictsofinterest{The author declares no conflicts of interest.}

\appendixstart
\appendix
\appendixtitles{yes}

\section{Details on Neutrino Emission and Proton--Photon Interaction Formalism}
\label{details}

This appendix outlines the Doppler transformation of photon fields in the jet comoving~\mbox{frame}.


\subsection{Doppler Transformations}
Quantities computed in the comoving frame are transformed to the observer frame using the Doppler factor:
\begin{equation}
\delta = \frac{1}{\Gamma\,(1-\beta\cos\theta_{\rm obs})},
\end{equation}
where $\Gamma$ is the bulk Lorentz factor, $\beta=v/c$, and $\theta_{\rm obs}$ is the viewing angle.

The angle between the local velocity vector $\mathbf{v}$ and the line-of-sight unit vector
$\mathbf{l}$ is:
\begin{equation}
\cos\theta = \frac{\mathbf{v}\cdot\mathbf{l}}{|\mathbf{v}||\mathbf{l}|}.
\end{equation}

The LOS vector components are:
\begin{equation}
l_x = \cos\phi_2 \cos\phi_1,\quad
l_y = \cos\phi_2 \sin\phi_1,\quad
l_z = \sin\phi_2.
\end{equation}

Photon energies transform as $E = \delta\,E'$ and intensities as appropriate powers of $\delta$
depending on the specific observable discussed in the main text.

\subsection{Line-of-Sight Geometry}
The LOS unit vector $\mathbf{l}$ is parameterized as follows:
\[
l_x = \cos\phi_2\cos\phi_1,\qquad l_y = \cos\phi_2\sin\phi_1,\qquad l_z = \sin\phi_2.
\]
The velocity vector is reversed in sign when calculating the Doppler factor for photons incoming into the jet comoving frame.

\section{Instrumental Sensitivities and Detection Criteria}
\label{app:instr_sens}

A note on sensitivity curves and detection thresholds adopted for the instruments considered in the detectability comparison. The plotted curves/bands are taken from the
corresponding experimental performance studies. A background-limited discovery claim would require atmospheric-neutrino background rates and analysis choices; this is beyond the scope of the present work and is noted explicitly in Section~\ref{results}.

\section{Normalization of the Model Neutrino Spectrum}
\label{app:norm}

\subsection{Source-Level Normalization}

The neutrino spectra presented in Figure~\ref{sed_plot} are now interpreted as differential neutrino power distributions per logarithmic energy interval:
\begin{equation}
\mathcal{P}_\nu(E) \equiv E^2\,\frac{d\dot N_\nu}{dE},
\end{equation}
with units of energy per unit time; in Figure~\ref{sed_plot}, it is plotted in $\mathrm{GeV\,s^{-1}}$.
For the normalization used in Figure~\ref{sed_plot}, we take the jet kinetic power from Table~\ref{Table-for-run-data}, \(L_k=2\times10^{38}\,\mathrm{erg\,s^{-1}}\), and assume that a fraction \(\eta_\nu=0.01\) is converted into neutrinos. Thus,
\begin{equation}
L_\nu = \eta_\nu L_k = 2\times10^{36}\,\mathrm{erg\,s^{-1}} .
\end{equation}
The model spectrum is normalized by enforcing
\begin{equation}
\int_{E_{\min}}^{E_{\max}} \mathcal{P}_\nu(E)\,\frac{dE}{E} = L_\nu,
\end{equation}
i.e., the integral over $d\ln E$ equals the assumed neutrino power. When plotted in GeV units, the conversion $1~\mathrm{erg}=624~\mathrm{GeV}$ is used.

\subsection{Flux at Earth and Comparison Units}

For a source at distance $d$, the corresponding spectral energy flux is:
\begin{equation}
E^2\,\frac{d\Phi_\nu}{dE} = \frac{\mathcal{P}_\nu(E)}{4\pi d^2},
\end{equation}
which preserves the normalization in physical units.

In Figure~\ref{fig:norm}, no additional scaling factor is applied beyond the geometric dilution by $4\pi d^2$. The plotted model curves therefore retain the normalization of Figure~\ref{sed_plot}, converted to flux at Earth.

\subsection{Case Used}
Unless otherwise stated, the normalization corresponds to the proton--photon model case~(b) discussed in the main text.


\begin{adjustwidth}{-\extralength}{0cm}
\reftitle{References}


\PublishersNote{}
\end{adjustwidth}

\end{document}